\documentclass[letterpaper,10pt]{article} 

\usepackage{opticameet3} 

\usepackage{amsmath,amssymb}
\usepackage[colorlinks=true,bookmarks=false,citecolor=blue,urlcolor=blue]{hyperref} 
\usepackage{caption}

\begin{document}

\title{Digital Twin of a Network and Operating Environment Using Augmented Reality}
\vspace{-10pt}
\author{
    Haoshuo~Chen\textsuperscript{(1)}, 
    Xiaonan~Xu\textsuperscript{(3)},    
    Jesse~E.~Simsarian\textsuperscript{(1)},  
    Mijail Szczerban\textsuperscript{(1)},  \\
    Rob Harby\textsuperscript{(2)},  
    Roland~Ryf\textsuperscript{(1)},  
    Mikael~Mazur\textsuperscript{(1)}, 
    Lauren~Dallachiesa\textsuperscript{(1)}, 
    Nicolas~K.~Fontaine\textsuperscript{(1)}, 
    John Cloonan\textsuperscript{(2)}, 
    Jim Sandoz\textsuperscript{(2)},
    David~T.~Neilson\textsuperscript{(1)}
}

\address{
    \textsuperscript{(1)} Nokia Bell Labs, 600 Mountain Ave., Murray Hill, NJ 07974, USA \\
    \textsuperscript{(2)} Nokia, 600 Mountain Ave., Murray Hill, NJ 07974, USA \\
    \textsuperscript{(3)} Aberdeen, NJ 07747, USA \\
}

\email{haoshuo.chen@nokia-bell-labs.com}
\vspace{-15pt}
\begin{abstract}
We demonstrate the digital twin of a network, network elements, and operating environment using machine learning. We achieve network card failure localization and remote collaboration over 86 km of fiber using augmented reality.
\vspace{-15pt}
\end{abstract}

\section{Introduction}
\vspace{-4pt}
A network digital twin is a simulation model of a communication system and its operating environment that enables applications such as the monitoring of network operations in real time, predictive maintenance, and testing ``what if'' scenarios before implementation on a production network. 
Recent work on digital twins of optical networks has developed accurate simulation and machine-learning (ML) models of the fiber transmission system~\cite{zhuge}. 
There has been an increased awareness that communication networks are physical systems that interact with and can be used to sense the environment~\cite{verizon}, motivating the digital-twin model to include the operating environment. Adding the physical and environmental information has multiple benefits, including improved physical connectivity visibility, a better understanding of shared risk groups, and a better facility security analysis.

In this work, we demonstrate an optical network digital twin model based on a graph neural network (GNN)~\cite{graph} with novel capabilities enabled by models of the physical network elements themselves as well as the operating environment. 
The network operators interact with the digital twin in real time using a distributed augmented reality (AR) application empowered with ML through remote computing. 
The AR application relies on a low-latency connection to a remote edge server that performs multiple computational functions. 
By using a 3-dimensional (3D) map of the network surroundings, 3D models of the network elements, and fault localization on the optical network, we show that the digital twin enables automated guidance of an on-site operator to a network element with a failure condition. 
When the operator views the network element with AR, a distributed ML-based image classification algorithm indicates the card that has the root-cause alarm condition. 
Finally, the AR application allows a real-time interactive collaboration session with a second operator remotely connected to a node after 86 km of fiber propagation so that knowledge can be shared across central and dispersed locations. 
Within the AR session, both operators can manipulate 3D computer-automated design (CAD) models of the network element and card as virtual 3D holograms, thereby enabling collaborative maintenance operations inside a metaverse \cite{schafer}. 

\begin{figure}[!b]
\vspace{-20pt}
  \centering
  \includegraphics[width=16.0cm]{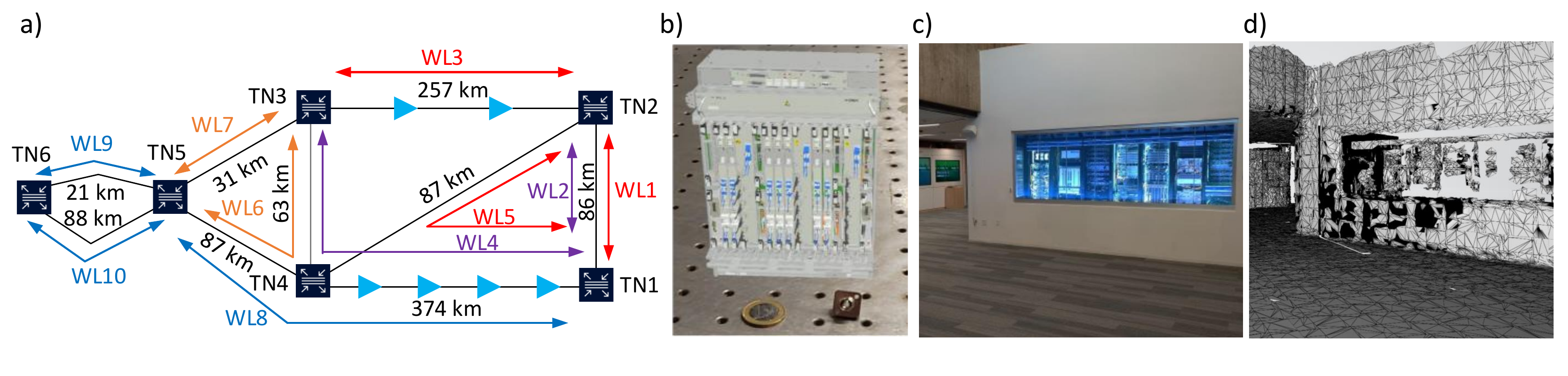}
  \vspace{-26pt}
\caption{a) Optical network topology, b) 3D digital model of a network element, c) image of the network environment and d) 3D mapping of the network environment.}
\label{fig1}
\end{figure}

\vspace{-4pt}
\section{Digital Twin of the Network and Its Operating Environment}
\vspace{-4pt}
We construct a digital twin representation of an optical transport network and the laboratory environment, encompassing the network topology, network equipment 3D models, and a 3D map of the environment. Fig.~\ref{fig1}a is a diagram of the optical network consisting of six commercial Nokia 1830 PSS optical transport nodes (TNs) with flexgrid reconfigurable optical add-drop multiplexers (ROADMs). Wavelength (WL) paths and fiber lengths are indicated in the figure. 
Fig.~\ref{fig1}b shows an AR image of a computer-generated downsized virtual hologram of one of the PSS-32 shelves of node TN1. 
The hologram was generated from 3D CAD models of the network element. 
Fig.~\ref{fig1}c is a photograph of the surrounding environment of the network, and Fig.~\ref{fig1}d is a 3D map of the same location created with a Microsoft HoloLens 2 AR headset (ARH)~\cite{AR}.


\begin{figure}[!t]
  \centering
  \includegraphics[width=16.2cm]{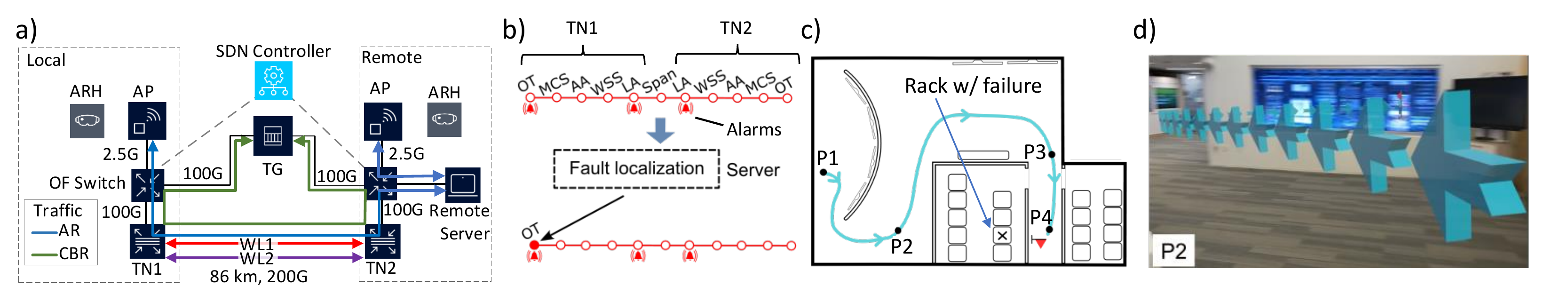}

\caption{a) Experiment setup for remote AR collaboration,
b) illustration of ML-based fault localization,
c) top-view 2D map of the network surroundings with virtual navigation markers,
and d) image of navigation guidance taken directly through the display of the local (on-site) ARH at the P2 location.}
\label{fig2}
\vspace{-25pt}
\end{figure}

Figure~\ref{fig2}a shows a diagram of the equipment used for the AR remote collaboration experiment. The optical transport nodes TN1 and TN2 are the same nodes as shown in the optical network diagram of Fig.~\ref{fig1}a and wavelength WL1 is used to carry traffic between the nodes for the experiment. WL1 has a line rate of 200~Gbit/s using 8-QAM modulation format and a low bit-error ratio in both directions. Two Centec V586 Openflow (OF) version 1.3 switches with 10G and 100G interfaces connect to 100G client interfaces on D5X500 flexible bitrate transponders at TN1 and TN2 that generate and terminate WL1. An instance of the Open Network Operating System (ONOS)~\cite{onos} software-defined network (SDN) controller controls the OF switches. 
Wi-Fi access points (APs) connect the local and remote AR headsets to the network with 2.5G connections to the OF switches at the local and remote sites, respectively. 
A 100G path through the network carries both the AR traffic from the local ARH to a remote server as well as a constant bitrate (CBR) stream from a 100G interface of a Spirent SPT-N12U traffic generator (TG) that produces bi-directional competition to the AR traffic.  

On-demand computing is enabled using a remote server that is equipped with an NVIDIA RTX A6000 graphics processing unit. 
The server assigns tasks to clients with AR capabilities, processes the requests from the clients through ML models, and synchronizes the operations between the clients.  
In the experiment, WL2 experiences frame losses and three severe network card alarms, as illustrated in Fig.~\ref{fig2}b.
The figure illustrates the various card-level network elements such as optical transponder (OT), line amplifier (LA), wavelength selective switch (WSS), array amplifier (AA), multicast switch (MCS), and fiber span.
A fault localization model~\cite{graph}, which employs GNNs and natural language processing, is executed on the remote server and successfully identifies the optical transponder of WL2 as the source of the failure by collecting the alarms from the network elements and utilizing the network connectivity graph, as illustrated in Fig.~\ref{fig2}b.
The AR applications were developed using Unity, OpenXR, and mixed reality toolkit~\cite{unity}. 
An on-site (local) operator wearing an ARH uses a hand-operated menu to choose between applications, set up parameters such as the server IP address, and connect or disconnect from the remote server.

\vspace{-4pt}
\subsection{Lab navigation}
\vspace{-1pt}
We developed an AR-based lab navigation application to assist network operators in efficiently navigating to their desired destination, e.g., a rack containing a failed network card.
A top view of the 2D lab map is presented in Fig.~\ref{fig2}c, where a path from the starting point (P1) to the rack containing the failed transponder (P4) is indicated by a series of blue virtual directional arrows and the destination is marked by a virtual red flag.
The on-site ARH operator receives the coordinates of the arrows and the flag from the remote server, which calculates the path using the A-star path-finding algorithm~\cite{astar} based on the environment's 3D map of Fig.~\ref{fig1}d.
Each rack has two network equipment shelves, and the height of the flag serves as an indication of the targeted shelf level. 
Fig.~\ref{fig2}(d) shows an AR navigation image captured directly through the display of the local ARH at the P2 location along the navigation path.

\vspace{-4pt}
\subsection{Network card identification}
\vspace{-1pt}
After locating the shelf, the next step is to indicate to the operator the network cards with alarms and identify the root-cause of failure.
Fig.~\ref{fig3}a shows a diagram of the process for network card identification and indication to the operator through ML-based computation at the remote server.
The computation uses the object detection model YoloV7~\cite{yolo}, which was fine-tuned using 824 captured images in order to detect and classify 11 different types of network cards and 2 different shelves. 
We employ the trainable bag-of-freebies method~\cite{bags} to improve accuracy and reduce the size of the training dataset.
The webcam of the ARH captures images at a rate of 5 frames per second, which are sent to the remote server. 
YoloV7 outputs the name of the detected card, a bounding box that encompasses the card, and a confidence score. 
A higher confidence score indicates the likelihood that the box contains the object, with a maximum score of 1.
The coordinates of the bounding boxes provide the relative positions of the detected cards.
Using the card arrangement information retrieved from the network element database, cards with alarms can be determined.
This information is then sent back to the ARH with additional color coding to indicate the nature of the alarm. 
Cards with a failure requiring replacement are indicated in red, while cards with an alarm that are not the main source of failure are indicated in blue. 
The bounding boxes, along with the card model names and confidence scores, are displayed on the ARH.
In the case of the shelf in Fig.~\ref{fig3}a, the flexible-bitrate OT card (model: D5X500Q) is determined to be the main source of failure with a confidence score of 85\%, and the LA card (model: ASWG) is also detected as having an alarm, but the GNN ML model of the network determines that it is not the root cause of the alarm. 
Note that the shelf image in Fig.~\ref{fig3}a contains four of the same LAs that are not indicated as faulty by the remote server computation.
Due to the knowledge of card positions, the LA with the alarm, which is rightmost among the four, has been successfully identified.

We tested the robustness of the remote classification computation to network congestion by introducing CBR competition to the AR traffic with the traffic generator shown in Fig~2a. We use the OpenFlow~\cite{openflow} meter table to prioritize the AR traffic, which restricts the maximum rate of the competing traffic to 90 Gb/s of the 100 Gb/s total. 
Fig.~\ref{fig3}b shows the maximum bitrate between the on-site ARH and the remote edge server and the total round-trip latency for the identification result, which is caused by the ML inference time of YoloV7 on the server and the network round-trip time over 86 km of single-mode fiber and the Wi-Fi link.
With prioritization of the AR traffic, we measured a maximum of 330 Mb/s bitrate with iPerf~\cite{iperf}, and $<$35 ms total round-trip latency for the card identification result.
Fig.~\ref{fig3}c shows histograms of the ML inference time and network round-trip data transfer time.

\vspace{-4pt}
\subsection{Interactive remote collaboration}
\begin{figure}[!t]
  \centering
  \includegraphics[width=16.0cm]{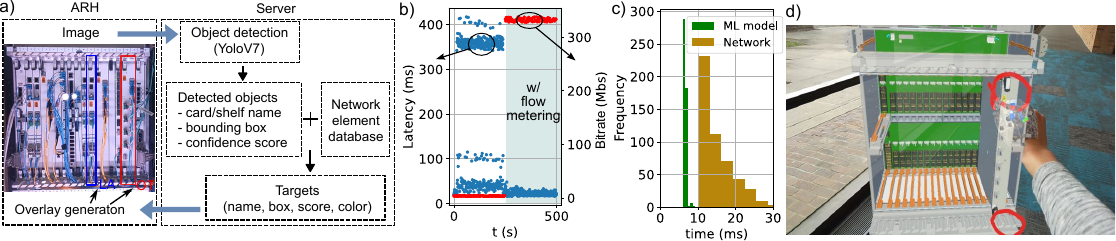}
  \vspace{-20pt}
\caption{a) Diagram of the process for network card identification using remote edge computing, 
b) maximum bitrate and object identification round-trip latency with and without CBR traffic metering,
c) histogram of ML inference and network data round-trip times for object identification with CBR metering, d) image of AR-assisted remote collaboration for card replacement}
\label{fig3}
\vspace{-18pt}
\end{figure}

\vspace{-1pt}
We developed a remote collaboration application so that a local network operator can collaborate in real time with a remote expert. 
For example, the local operator can receive guidance for the process of replacing the network card by manually manipulating the virtual 3D models of the transport system cards and shelf that are generated as 3D images in the virtual environments of both the operator and remote expert.
The remote edge server synchronizes the position and orientation of the digital 3D models for both participants. 
Fig.~\ref{fig3}d shows an AR image in which the expert utilizes mid-air drawing of red circles to indicate the locations of the two latches that must be released prior to removing the card and demonstrates the card-replacement procedure to the local operator. The application also supports real-time voice and video communication between the two headsets.

\vspace{-6pt}
\section{Conclusion}
\vspace{-4pt}
In this work, we demonstrate a digital twin of a network, network elements, and operating environment utilizing ML and remote edge computation. Through interaction in an AR virtual environment, the digital twin enabled indoor navigation, network card failure identification and localization, and remote collaboration over an 86-km optical link. These innovations demonstrate the potential of AR and ML in network management and maintenance.

Supplemental video: https://youtu.be/RJMDRjCIBFI

\end{document}